\documentclass[12pt]{article}
\usepackage{natbib,graphicx,setspace,lscape,longtable,epstopdf,xr}
\usepackage{mathrsfs,amsmath,amsthm,amssymb,color}
\usepackage{natbib,epsfig,graphicx,pdfpages}
\usepackage{caption,multirow,tabularx,booktabs,siunitx}
\usepackage{comment}
\usepackage{enumitem}
\usepackage{rotating}
\usepackage{float}
\usepackage{bm}
\usepackage{ulem}
\usepackage{CJK}
\usepackage{mathtools}
\usepackage{ragged2e}
\usepackage{environ}
\usepackage{dcolumn}
\usepackage{caption}
\usepackage{subcaption}
\usepackage{siunitx,booktabs}
\usepackage{algorithm}
\usepackage{algorithmic}

\bibpunct{(}{)}{;}{a}{,}{,}

\newtheorem{theorem}{Theorem}
\newtheorem{lemma}{Lemma}
\newtheorem{proposition}{Proposition}
\newtheorem{definition}{Definition}
\newtheorem{assumption}{Assumption}

\newtheorem{example}{Example}
\newcommand{\csection}[1]
    {\begin{center}
        \stepcounter{section}
        {\bf\large\arabic{section}. #1}
    \end{center}
}

\newcommand{\csubsection}[1]{
\begin{center}
\stepcounter{subsection}
{\it\arabic{section}.\arabic{subsection}. #1}
\end{center}
}

\def\beq{\begin{equation}}
\def\eeq{\end{equation}}
\def\beqr{\begin{eqnarray}}
\def\eeqr{\end{eqnarray}}
\def\beqrs{\begin{eqnarray*}}
\def\eeqrs{\end{eqnarray*}}
\def\bet{\begin{theorem}}
\def\eet{\end{theorem}}
\def\bel{\begin{lemma}}
\def\eel{\end{lemma}}
\def\bep{\begin{proposition}}
\def\eep{\end{proposition}}
\def\bep{\begin{assumption}}
\def\eep{\end{assumption}}
\def\bg{\begin{figure}[tbph]\begin{center}}
\def\eg{\end{center}\end{figure}}

\def\bc{\begin{center}}
\def\ec{\end{center}}
\newtheorem{remark}{Remark}

\def\wt{\widetilde}
\def\wh{\widehat}

\def\mA{\mathcal{A}}
\def\mD{\mathcal{D}}
\def\mK{\mathcal{K}}
\def\mL{\mathcal{L}}
\def\mM{\mathcal{M}}
\def\mR{\mathbb{R}}
\def\mE{\mathbb{E}}

\def\bI{\mathbb{I}}
\def\one{\bm 1}
\def\argmax{\mbox{argmax}}
\def\diag{\mbox{diag}}

\def\mA{\mathcal{A}}
\def\mS{\mathcal{S}}
\def\mU{\mathcal{U}}

\usepackage{hyperref}
 \hypersetup{
	colorlinks=true,
	filecolor=blue,
	citecolor = blue,
	urlcolor=cyan,
}

\setlist[enumerate]{label*=(A\arabic*)} 

\newcolumntype{C}{>{\centering\arraybackslash}X}

\numberwithin{equation}{section}

\begin{document}
\begin{CJK}{GBK}{song}
\begin{center}
{\bf\Large {Subsampling Spectral Clustering for Stochastic Block Models in Large-Scale Networks}}\\
\bigskip

Jiayi Deng$^{1,2}$, Danyang Huang$^{1,2*}$, Yi Ding$^{1,2}$, Yingqiu Zhu$^{3}$, Bingyi Jing$^4$, and Bo Zhang$^{1,2*}$

{\it\small
$^1$ Center for Applied Statistics, Renmin University of China, Beijing, China;\\
$^2$ School of Statistics, Renmin University of China, Beijing, China;\\
$^3$ School of Statistics, University of International Business and Economics;\\
$^4$ Department of Statistics and Data Science, Southern University of Science and Technology.
}

\end{center}

\begin{footnotetext}[1]
{ Danyang Huang and Bo Zhang are co-corresponding authors. They have contributed equally to this paper. Danyang Huang, Center for Applied Statistics, School of Statistics, Renmin University of China at Beijing, 59 Zhongguancun Street, 100872, China; dyhuang@ruc.edu.cn. Bo Zhang, Center for Applied Statistics, School of Statistics, Renmin University of China at Beijing, 59 Zhongguancun Street, 100872, China; mabzhang@ruc.edu.cn. This work was supported by the
National Natural Science Foundation of China (grant numbers 12071477,
11701560, 71873137); fund for building world-class universities (disciplines) of Renmin University of China. The authors gratefully acknowledge the support of Public Computing Cloud, Renmin University of China.}
\end{footnotetext}

\begin{singlespace}
\begin{abstract}
 The rapid development of science and technology has generated large amounts of network data, leading to significant computational challenges for network community detection. Here, we propose a novel subsampling spectral clustering algorithm to identify community structures in large-scale networks with limited computing resources. More precisely, we first construct a subnetwork by simple random subsampling from entire network, and then we extend the existing spectral clustering to the subnetwork for estimating the community labels for entire network nodes. As a result, for large-scale datasets, the method can be realized even using a personal computer. Moreover, under the stochastic block model and its extension, the degree-corrected stochastic block model, the theoretical properties of the subsampling spectral clustering method are correspondingly established. Finally, to illustrate and evaluate the proposed method, a number of simulation studies and two real data analyses are conducted.\\

\end{abstract}
\end{singlespace}

\newpage
\csection{INTRODUCTION}

Community detection is an important research direction in network analysis \citep{newman2004finding, fortunato2010community}, and it aims to cluster the nodes into different groups with high edge concentrations within the same cluster and low concentrations between different ones \citep{girvan2002community, lancichinetti2009community}. Network community detection is widely applied in various research areas, including, but not being limited to, computer science \citep{agarwal2005beyond, tron2007benchmark}, social science \citep{zhao2011community, lee2017time}, and biology \citep{rives2003modular, chen2006detecting, nepusz2012detecting}. Recently, the advances of science and technology created large amounts of online network data. However, even if the computing techniques have improved significantly, directly dealing with large-scale network data remains challenging, especially when available computing resources are limited \citep{harenberg2014community, wang2018optimal, wang2019information}.

\begin{figure}
 \centering
     \begin{subfigure}[b]{0.46\textwidth}
         \centering
         \includegraphics[width=\textwidth]{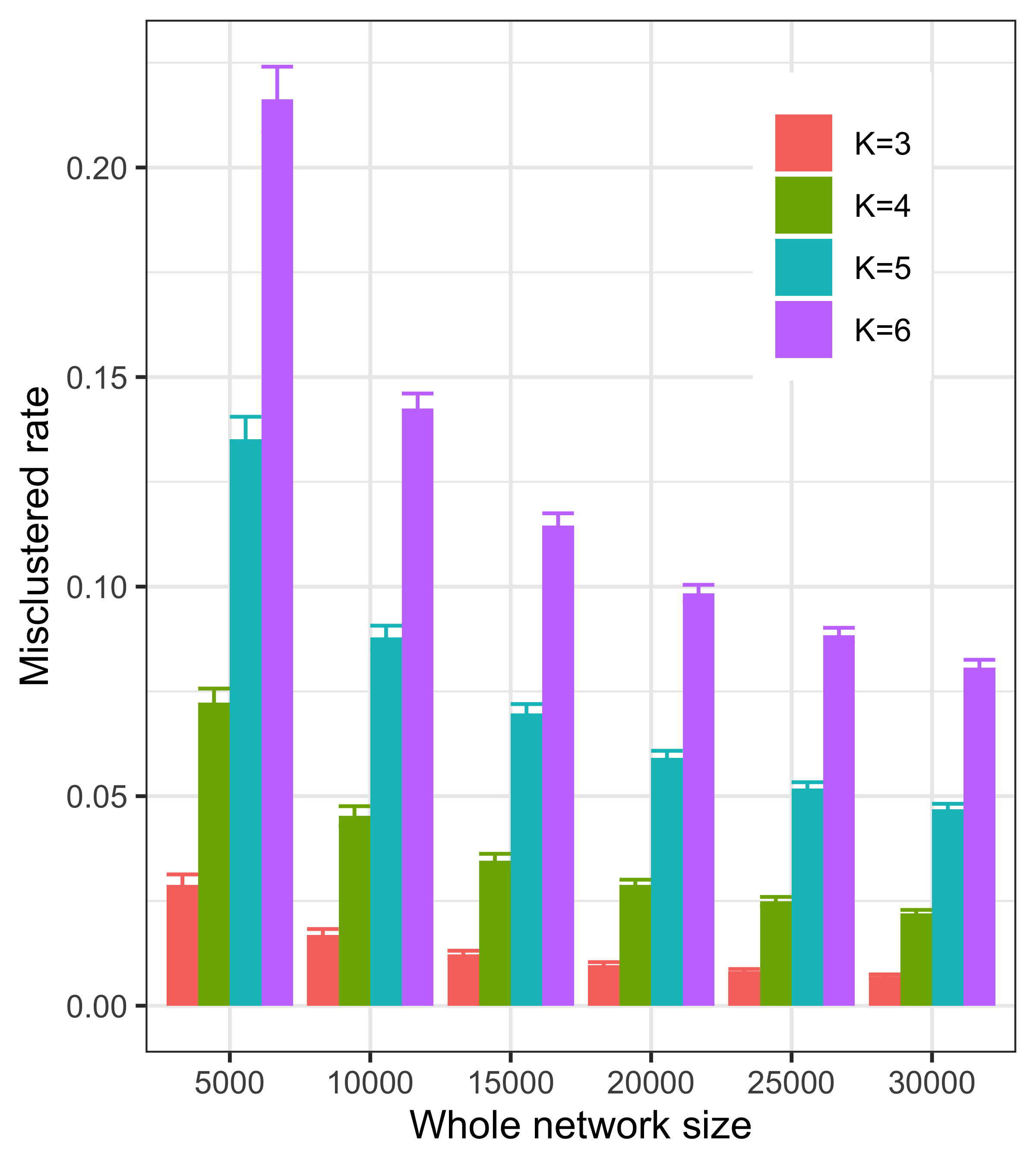}
     \end{subfigure}
     \hfill
     \begin{subfigure}[b]{0.46\textwidth}
         \centering
         \includegraphics[width=\textwidth]{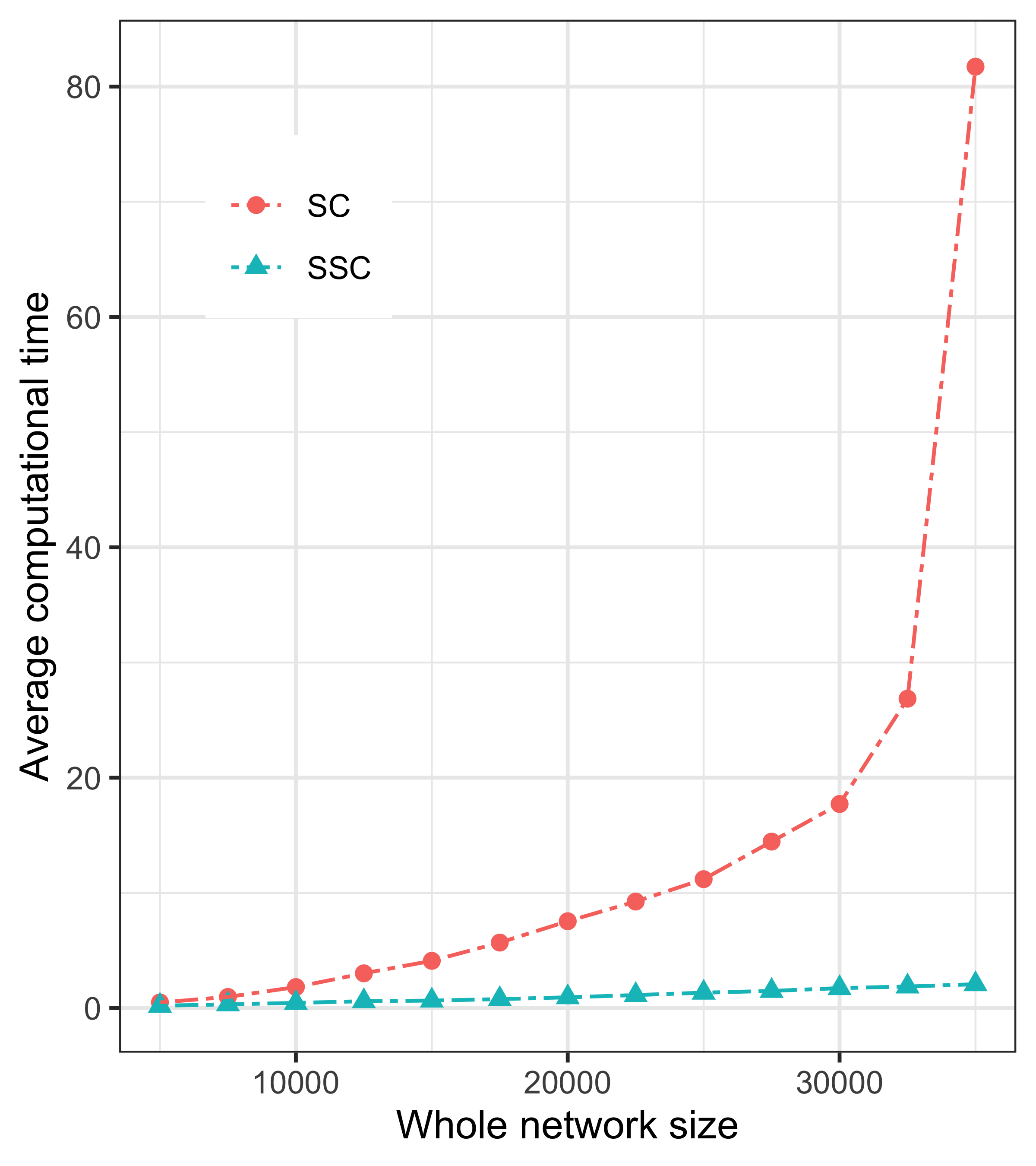}
     \end{subfigure}
   \caption{\small  The misclustered rate (left panel) versus whole network size $N$. Here, the subsample size is $n=\lceil 15\{\log(N)\}^{2}\rceil$. Further, the standard deviation is reported by the error bar in the left panel. The average computational time (right panel) of the SSC and SC are further compared as $N$ increases from 5,000 to 35,000.}\label{fig: consistency}
\end{figure}

In community detection literature, spectral clustering is one of the most popular methods because it is easy to implement and has well-founded theoretical guarantees \citep{ ng2002spectral, rohe2011spectral, krzakala2013spectral, binkiewicz2017covariate}. Given a network with $N$ nodes and $K$ different clusters, spectral clustering embeds each node in a $K$-dimensional space based on singular value decomposition (SVD). The computational complexity of spectral clustering is $O(N^{3})$ if a full SVD is performed \citep{yan2009fast, li2011time, chen2011large}, which is difficult to afford for analyzing large-scale networks. To speed up the spectral clustering, many fast SVD algorithms based on randomization techniques have been proposed, including \cite{halko2011finding}, \cite{feng2018faster}, and \cite{martin2018fast}. The computational cost of these methods is at the order of $O(N^{2})$. To be more specific, as shown in the right panel of Figure \ref{fig: consistency}, when the network size is $N=35,000$, running spectral clustering with a fast SVD algorithm takes 81.7 seconds. Moreover, as $N$ grows, the computational cost increases dramatically. Therefore, existing spectral clustering methods are still intractable for many real-world applications involving large-scale datasets.

To deal with this computational challenge brought by large-scale datasets, subsampling is a valuable tool \citep{politis1999subsampling}. Its advantage is that we could obtain a computationally efficient and consistent estimator based on a small subsample \citep{wang2019information, wang2021optimal, yu2022optimal}. Although subsampling pays the price of statistical convergence, it makes the existing methods feasible in large-scale data analysis. In the literature, various sampling designs have been proposed to derive representative samples of a given network, which include node sampling methods \citep{snijders1999non,bhattacharyya2015subsampling,lunde2019subsampling}, and edge sampling methods \citep{gonen2011counting, eden2017approximately, li2020network}.

Network subsampling strategy strategies have been widely used for solving the problem of image segmentation \citep{fowlkes2004spectral,wang2011approximate} and network community detection \citep{mukherjee2021two,zhang2020randomized}. Specifically, \cite{fowlkes2004spectral} proposed an approximate spectral clustering based on the Nystr{\"o}m method, which samples the columns of the similarity matrix and constructs a low-rank matrix completion to approximate the full matrix. Moreover, \cite{mukherjee2021two} developed two divide-and-conquer algorithms for identifying the community labels for large-scale networks in a distributed fashion. More recently, \cite{zhang2020randomized} adopted a simple sampling strategy to obtain a sparsified entire network. The main challenge of these methods is that to obtain the community labels of all nodes, we have to apply spectral clustering on an $N\times N$ low-rank matrix, which is still computationally intractable with limited computational resources. This motivates us to develop a new community detection method that can focus on a small subnetwork extracted by network subsampling.

\begin{figure}
  \centering
  \includegraphics[width=5.3 in]{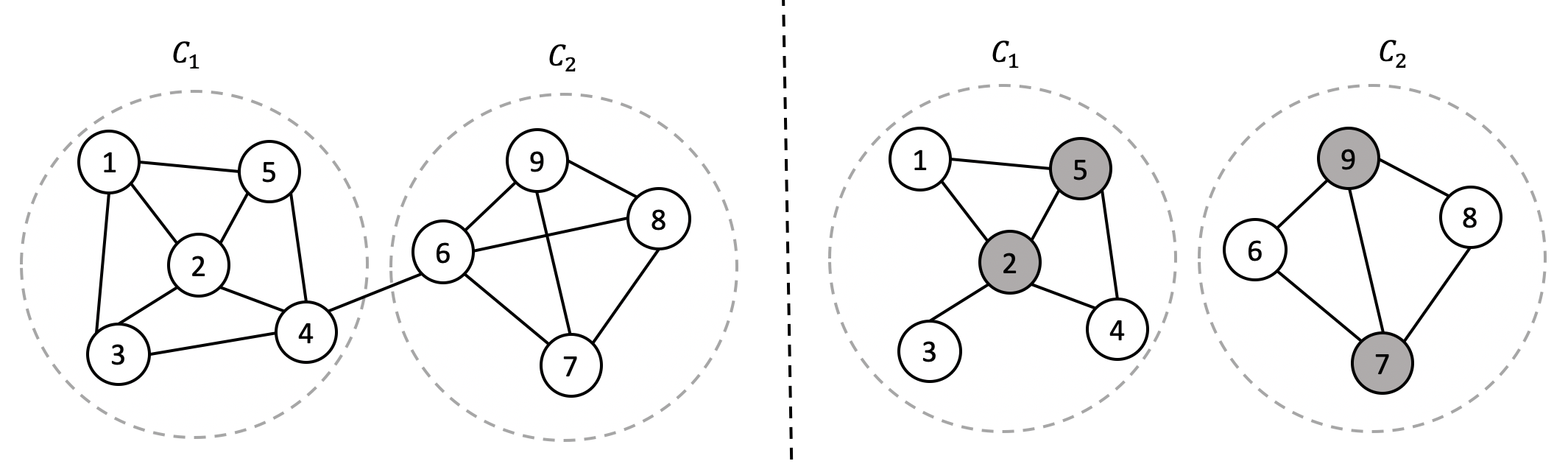}
  \caption{\small Example of network subsampling. A network with nine nodes and two communities ($C_1,C_2$) is displayed in the left panel. The right panel is a subnetwork, in which the selected nodes are gray, the edges connected by selected nodes are collected while the edges between unselected nodes are deleted.}\label{fig_demo}
\end{figure}

Specifically, we investigate selecting a small node set to extract the network structure information with limited computational cost. Considering large-scale networks, the network structure information is contained in the connections among nodes and can be represented by network adjacency or Laplacian matrices. However, such matrices are high-dimensional and could lead to high computational costs. To solve the problem, we propose selecting a small node set and then consider extracting the network community structure only through connections related to the selected nodes. To illustrate, we provide an example of network subsampling as shown in Figure \ref{fig_demo}, where the graph contains nine nodes assigned to two communities. It is remarkable that we can perfectly identify the community labels for all nodes based only on the connections related to the selected nodes. Moreover, because we only select a small subset of the total nodes, we can use a much lower dimensional matrix to represent the connection information of network nodes.

In this way, we propose a subsampling spectral clustering (SSC) method for large-scale networks under the constraint of limited computing resources. The SSC method estimates the community labels for entire networks by applying spectral clustering to a subnetwork, which is much smaller than the entire network. Specifically, we theoretically verify that the subsampling size $n$ can be as small as $\Omega\{(\log{N})^3\}$. Consequently, the computational complexity of the SSC method could be as low as $O\{N(\log{N})^3\}$. Moreover, we established the upper bound of the misclustered rate for the SSC method under the stochastic block model (SBM, \citealt{holland1983stochastic}) and degree-corrected SBM (DCSBM, \citealt{karrer2011stochastic}), respectively.

The remainder of the paper is organized as follows. In Section 2, we propose the SSC algorithm for large-scale networks. In Section 3, we establish the theoretical properties of the SSC algorithm for both the SBM and the DCSBM. The simulation and real data studies are presented in Section 4. Section 5 highlights the main conclusions and discusses future research. All technical proofs are presented in the Appendices of the Supplementary Material.

\csection{SUBSAMPLING SPECTRAL CLUSTERING FOR STOCHASTIC BLOCK MODELS}

Here, we first introduce the SBM, DCSBM and corresponding notations. Then, we propose a subsampling spectral clustering method to identify the community labels for large-scale networks generated from the SBM or DCSBM.

\csubsection{Stochastic Block Model and Its Extension}

The stochastic block model is an important random graph model for studying community detection \citep{holland1983stochastic, rohe2011spectral}. Under a SBM with $N$ nodes and $K$ communities, define a symmetric probability matrix $B=(B_{kk'}) \in (0,1)^{K\times K}$ and a label vector $z=(z_1,\cdots, z_N)^{\top}\in [K]^{N}$, where $[K]=\{1,\cdots, K\}$. Then, its adjacency matrix $A=(A_{ij})\in \{0,1\}^{N\times N}$ is assumed to be symmetric with zero diagonals and, for all $i>j$, $A_{ij}=A_{ji} \sim \rm{Bernoulli}(B_{z_iz_j})$ independently. It is noteworthy that, for any $i$, $j\in [N]$, the probability of an edge between node $i$ and $j$ depends only on their community memberships.

The DCSBM \citep{karrer2011stochastic} is generalized from the SBM, which introduces node-specific parameters to allow for degree heterogeneity within communities. Specifically, given parameters $(z, B)$, the probability of an edge between nodes $i$, $j$ is represented by $ P(A_{ij}=1)= \theta_i B_{z_iz_j} \theta_j$, where the parameter $\theta_i$ characterizes the individual activeness of node $i$. To ensure the identifiability of this model, we assume that $\sum_{i: z_i=k} \theta_i=1$ for all $k=1,\cdots, K.$ Let $N_{k}=\sum_{i=1}^{N}\bI(z_{i}=k)$ denote the size of the $k$th community, where $\bI(\cdot)$ is an indicator function.

Throughout this work, we investigate to identify the community labels for assortative networks \citep{amini2018semidefinite}. Namely, we assume that $\max_{k\neq l} B_{kl}< \min_{k}B_{kk}$. Moreover, in the following discussion, we consider the number of communities $K$ as a fixed constant.

\csubsection{Subsampling Spectral Clustering}

We propose the SSC method for a large-scale network generated from the SBM or the DCSBM. Owing to computational complexity considerations, we first subsample a subset of nodes $\mS$ from the entire network and then identify network communities using a spectral method to the normalized sub-adjacency matrix .

\begin{figure}
  \centering
  \includegraphics[width=5.3 in]{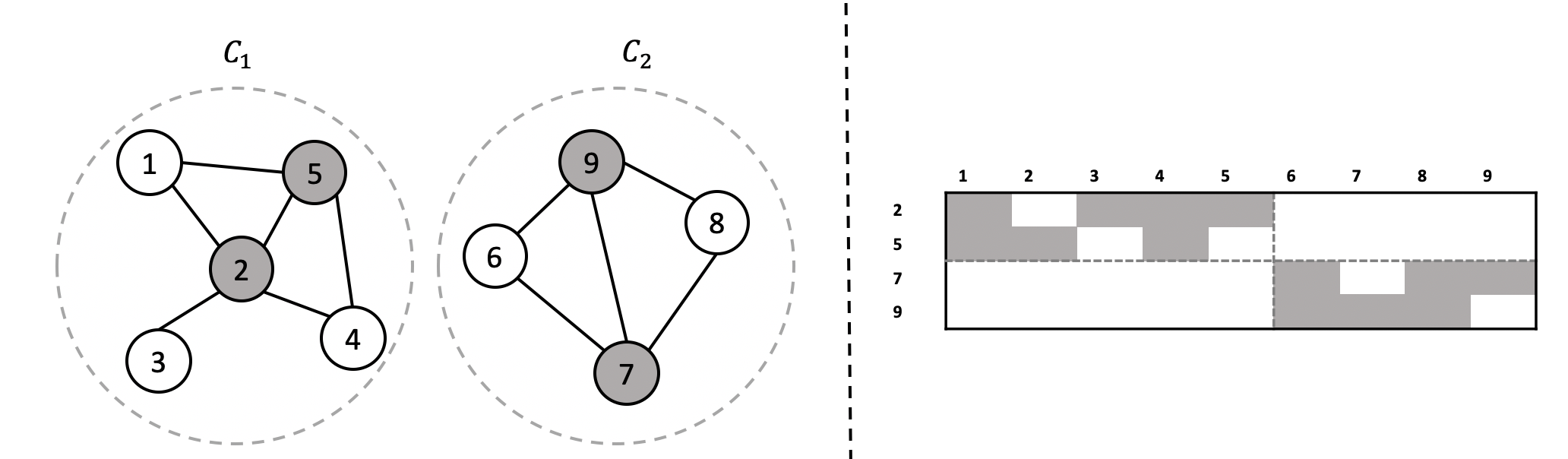}
  \caption{\small  An illustration of a sub-adjacency matrix. The left panel displays a network with nine nodes, two communities ($C_1, C_2$) and four selected nodes indicated by gray. The transpose of sub-adjacency matrix $(A^{s})^{\top}$ is shown in right panel represents, where the gray scale indicates that there is an edge between the corresponding pair of nodes.}\label{fig_sub_adj}
\end{figure}

Specifically, for a network with $N$ nodes, we collect $n$ nodes from entire networks via simple random sampling. Simple random sampling is a uniform subsampling method, where the subsampling distribution is $p_{i}=n/N$ for $i=1,2,\cdots, N$. Let $\mS=\{s_j: s_j \in [N], 1\leq j\leq n\}$, where $ s_{j}$ denotes the selected node. Moreover, we define a sub-adjacency matrix $A^{s}=( A_{ij}^s)\in \mR^{N\times n}$, where $A_{ij}^s=A_{is_j}$ for $1\le i \le N$, $1\le j \le n$. The sub-adjacency matrix is used to represent the connections related to the selected node set $\mS$. For assortative networks, nodes within a community often have a similar connection intensity to the selected nodes, whereas nodes in different communities have a different connection intensity from the selected nodes. For instance, we display a network with nine nodes and two communities as shown in the left panel of Figure \ref{fig_sub_adj}, where the selected nodes are gray. The right panel shows the transpose of sub-adjacency matrix $(A^{s})^{\top}$. There are clear block structures in the sub-adjacency matrix, and nodes within a community are gathered in the same block. Therefore, based on the proper selected node set, the sub-adjacency matrix can contain almost all network community structure information.

Then, we introduce a definition for the normalized sub-adjacency matrix. Considering that, for adjacency matrix $A$, its normalized version can be denoted by $L=D^{-1/2}AD^{-1/2} \in \mR^{N\times N}$, where $D$ is a diagonal matrix with the $i$th diagonal element being $\sum_{j} A_{ij}, (1 \le i \le N)$. Then, a normalized sub-adjacency matrix can be defined as
 \beqr\label{equ_sublap}
L^{s} = (D^{s}_{r} )^{-1/2}A^{s} (D^{s}_{c} )^{-1/2} \in \mR^{N\times n},
 \eeqr
 where $D^{s}_{r}=\diag\{ (D^{s}_{r,i})_{i=1}^{N}\}$ and $D^{s}_{c}= \diag\{(D^{s}_{c,j})_{j=1}^{n}\}$ are defined as the out- and in-degree matrices of the subsampled node set $\mS$. The entries are $D^{s}_{r,i}=\sum_{j=1}^{n} A_{ij}^s$ and $D^{s}_{c,j}=\sum_{i=1}^{N}A_{ij}^s$. We next discuss how to estimate the underlying community labels $z$ based on the normalized sub-adjacency matrix. We discuss this separately in two different cases, i.e., SBM and DCSBM.

\textbf{Case 1 (SBM).} In the SBM framework, we apply spectral clustering to the normalized sub-adjacency matrix. Define a truncated SVD of $L^{s}$ as $L^{s}_{K}=\wh{U}\wh{\Sigma}\wh{V} ^{\top}\in \mR^{N\times n}$, where $\wh{\Sigma} \in \mR^{K \times K}$ is a diagonal matrix with the largest $K$ singular values of $L^{s}$ sorted in absolute decreasing order, $\wh{U}\in \mR^{N\times K}$ and $\wh{V}\in \mR^{n \times K}$ are the corresponding left and right singular vector matrices, respectively. We consider $\wh{U}$ as the approximated embedding vectors of all network nodes. Specifically, let $u_{i}=(\wh{U}_{i1},\cdots, \wh{U}_{iK})$, that is, $\wh{U}=(u_{1},\cdots, u_{N})^{\top}$, and then $u_i$ is considered as the embedding vector of node $i$. As a result, the community labels can be obtained by applying the k-means algorithm to the rows of $\wh{U}$.

\textbf{Case 2 (DCSBM).} In the DCSBM framework, we apply spherical spectral clustering \citep{lei2015consistency} to the normalized sub-adjacency matrix. Specifically, we let $v_{i}= u_i/\|u_i\|$ where $\| \cdot \|$ denotes the Euclidean norm of a vector, and we consider $v_i$ as the embedding vector of node $i$. Moreover, let $\wh{U}^{*}= (v_1,\cdots, v_N)^{\top}$ to be the row-normalized version of $\wh{U}$. Then, we perform k-means clustering to the rows of $\wh{U}^{*}$ to obtain the community labels for entire network nodes. As a result, the partition results are recorded as $\wh{C}_{k}=\{i: \hat{z}_{i}=k, 1\leq i\leq N\} (k=1,\cdots, K)$, where $\hat{z}_{i}$ is the estimate label of node $i$. For simplicity, the extension of a spectral clustering algorithm by subsampling is referred to as the SSC method. The SSC procedure for large-scale networks generated from the SBM or DCSBM is described in Algorithm \ref{alg_ssc}.

\begin{algorithm}[h]
\caption{Subsampling Spectral Clustering (SSC)}
\begin{algorithmic}
\STATE \textbf{Input}: adjacency matrix $A\in \mathbb{R}^{N\times N}$, number of communities $K$;\
\begin{itemize}
\item[1.] Collect $n$ nodes from entire node set $[N]$ by simple random subsampling and record the selected subset of nodes as $\mS$;
\item[2.] Form sub-adjacency matrix $A^{s} \in \mR^{N\times n}$ and compute its normalized sub-adjacency matrix , $L^{s} \in \mR^{N\times n}$, as defined in \eqref{equ_sublap}; \
\item[3.] Conduct a truncated SVD on $L^{s}$ and find its largest $K$ singular values and the corresponding left-singular vectors (i.e., $\wh{U}$);\\
\item[4.] (\textbf{SBM}) Conduct k-means to cluster the rows of $\wh{U}$ into $K$ clusters $\wh{C}_{1}, \cdots, \wh{C}_{K}$;\\
\item[$4'$.] (\textbf{DCSBM}) Calculate $\wh{U}^{*}$ by normalizing each row of $\wh{U}$ to have unit length, and conduct k-means to cluster rows of $\wh{U}^{*}$ into $K$ clusters $\wh{C}_{1}, \cdots, \wh{C}_{K}$;\\
\end{itemize}
\STATE \textbf{Output}: partition results $\wh{C}_{1}, \cdots, \wh{C}_{K}$.
\end{algorithmic}\label{alg_ssc}
\end{algorithm}

\begin{remark}[Determine the number of communities]\label{remark: determine} For real-world datasets with unknown number of communities, we adopt the corrected bayesian information criterion, proposed by \cite{hu2020corrected}, and combine the proposed method to estimate $K$. Specifically, let $\mK$ denote the candidate set for the number of communities, then we evaluate each candidate $K' \in \mK$ by following steps. First, based on the sub-adjacency matrix $A^{s}$, we apply the SSC method to $A^s$ to obtain the label estimates $\hat{z} \in [K']^{N}$. Then, under the SBM/DCSBM, we compute the log-likelihood function associated with $A^{s}$, $\hat{z}$, and denote this log-likelihood function as $\ell(K', \hat{z}, A^s)$. Lastly, according to \cite{hu2020corrected}, we calculate the corrected bayesian information criterion by $f(K',\hat{z},A^{s})=\ell(K', \hat{z}, A^s) -\left\{n\log{K'}+ \frac{K'(K'+1)}{4}\log{(Nn)}\right\}.$ As a result, the optimal solution is $\wh{K}=\argmax_{K'\in \mK} f(K', \hat{z},A^s).$
\end{remark}

\csubsection{Computational Complexity of the SSC Algorithm}

\begin{figure}[H]
   \centering
     \begin{subfigure}[b]{0.65\textwidth}
         \centering
         \includegraphics[width=\textwidth]{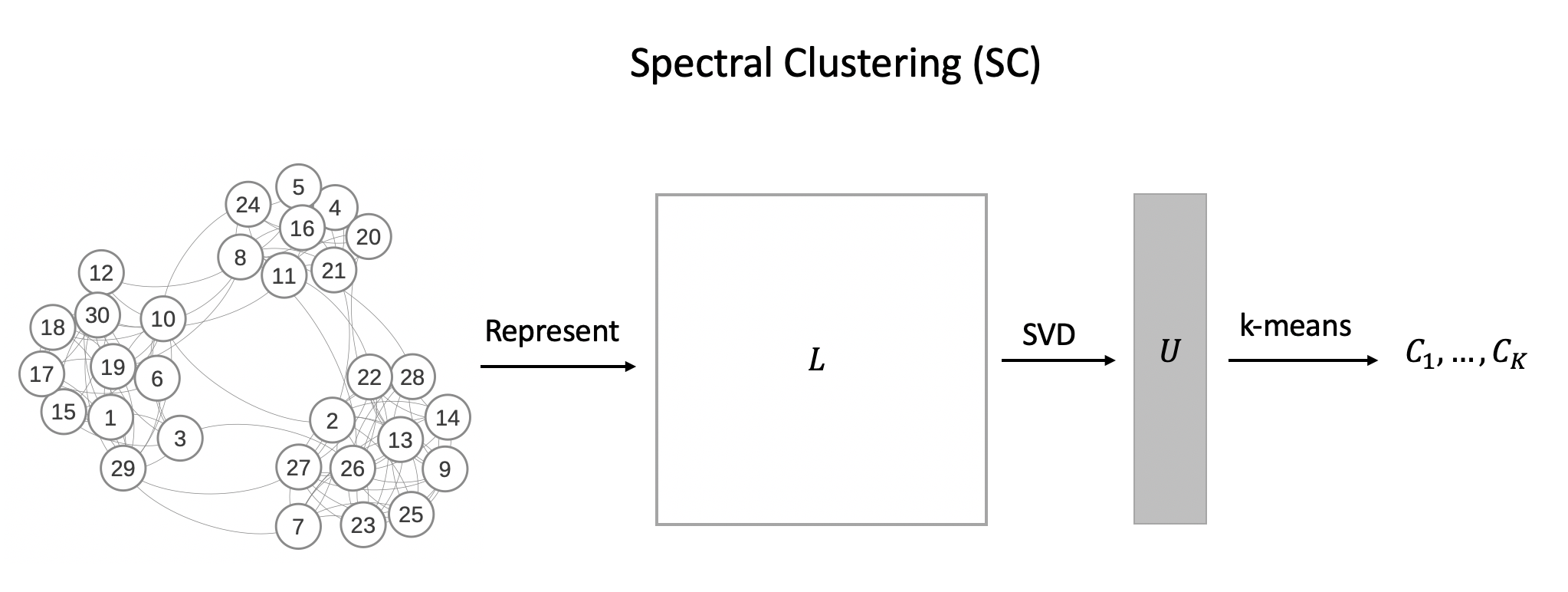}
     \end{subfigure}
     \begin{subfigure}[b]{0.65\textwidth}
         \centering
         \includegraphics[width=\textwidth]{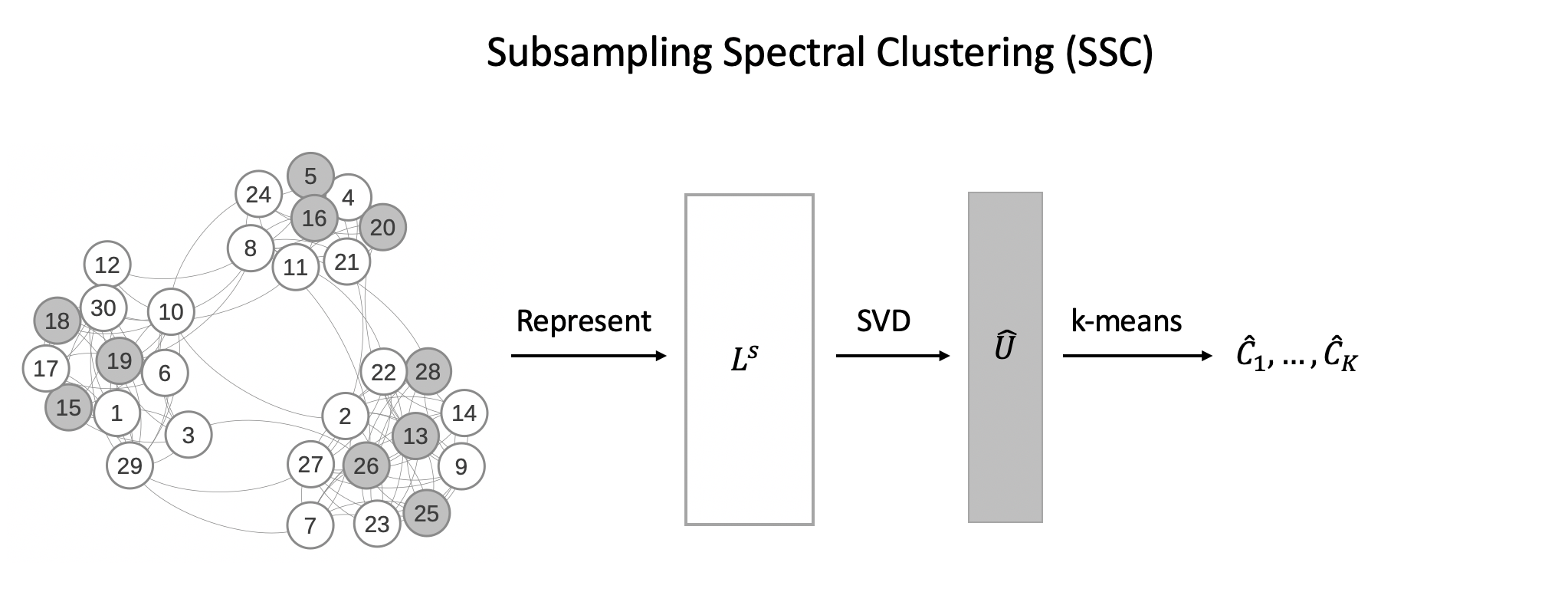}
     \end{subfigure}
  \caption{\small Comparison between spectral clustering and SSC under the SBM. The upper panel represents the spectral clustering algorithm, which applies the k-means algorithm to the rows of the largest $K$ eigenvectors of $L$ (i.e., $U$) to obtain the clustering results. The lower panel shows the SSC algorithm for SBM, which first applies subsampling to extract a subset of the entire node set where the gray nodes are the selected ones. Then, we form a normalized sub-adjacency matrix  $L^{s}$, and, through SVD, we find its largest $K$ left singular vectors, $\wh{U}$. We also apply the k-means algorithm to the rows of $\wh{U}$ to obtain the clustering results.}\label{fig_ssc}
\end{figure}

To illustrate the computational advantage of the proposed method, under the SBM, we compare SSC with the existing spectral clustering algorithm for the entire network in Figure \ref{fig_ssc}. Evidently, unlike the existing spectral clustering method, SSC creates a normalized sub-adjacency matrix  with a much lower column dimension based on the small subsample. In other words, SSC is conducted on a small part of the connections in the entire network. Consequently, its clustering results may not be as accurate as those obtained by spectral clustering. However, it can obtain cluster labels for the entire network with limited computing resources. This could make network clustering feasible even using a personal computer.

Moreover, we provide the following proposition to discuss the computational complexity of the SSC algorithm.
\begin{proposition}[\textbf{Computational complexity}]\label{pro_sscTime}
Suppose that the subset of nodes $\mS$ is collected by simple random sampling $n$ nodes from $[N]$. Then, for both the SBM and DCSBM, the computational complexity of community detection for the entire network, based on the SSC algorithm is $O(Nn)$.
\end{proposition}
\noindent We demonstrate Proposition \ref{pro_sscTime} via the following steps. First, the complexity of collecting the subset $\mS$ by simple random sampling from $[N]$ is $O(N)$ according to \cite{vitter1985random}. Second, forming the sub-adjacency matrix $A^{s}$ and computing $L^{s} \in \mR^{N\times n}$ each requires $O(Nn)$ computations. Third, the truncated SVD of $L^{s}$ costs $O(Nn)$ running times \citep{feng2018faster, martin2018fast}. Lastly, for SBM, conducting k-means to cluster the rows of $\wh{U}$ takes $O(N)$ \citep{hartigan1979algorithm}. While for DCSBM, calculating $\wh{U}^{*}$ and applying k-means to cluster the rows of $\wh{U}^{*}$ each requires $O(N)$ computational time, respectively. As a result, for the SBM/DCSBM, the overall computational complexity of the SSC algorithm is $O(Nn).$ Since the computational cost of existing spectral clustering is in the order of $O(N^2)$ and $n<<N$, the SSC method is more computationally efficient. Next, we discuss the theoretical properties of the SSC algorithm in detail.

\csection{THEORETICAL PROPERTIES OF THE SSC ALGORITHM}

In this section, we first analyze the clustering accuracy of the the SSC algorithm under the SBM via the following steps. First, we introduce some necessary conditions and subsequently discuss the required subsample size to ensure the effectiveness of the selected sample. Second, we establish the theoretical properties of the SSC algorithm based on population analysis. Third, we show the consistency of the largest $K$ left-singular vectors. Lastly, we provide the upper bound of the misclustered rate caused by the SSC algorithm. Furthermore, under the same assumptions, we discuss the consistency of the SSC algorithm under the DCSBMs.

\csubsection{Basic Assumptions and Required Subsample Size}

Before analyzing the theoretical properties of SSC, the following assumptions are considered.
\begin{enumerate}
\item(Balance Level) Let $N_{\rm min}= \min_{k}N_{k}$ denote the minimum community size and assume $N_{\rm min}= \alpha_{N} N$ where the balance level parameter such that $\alpha_{N}=\Omega\{(\log{N})^{-1}\};$\label{ass: balance}
\item(Network Sparsity) Assume the connectivity matrix $B=\rho_N\wt{B}$, where $\wt{B} \in (0,1)^{K\times K}$ is a constant matrix and $\rho_N \to 0$ at a rate of $\rho_NN/\log{N}\to \infty$.\label{ass: sparsity}
\end{enumerate}
\noindent
Assumption \ref{ass: balance} is introduced to specify the cluster size balance level, where $\Omega\{g(N)\}$ means that, for the set of all $f(N)$, there exist positive constants $a_0$ and $N_0$ such that $f(N)\ge a_0g(N)$ for all $N \ge N_0$ \citep{knuth1976big}. This assumption allows the sample sizes in different clusters to be of different orders to some extent, which is also discussed by \cite{lei2015consistency}.
Assumption \ref{ass: sparsity} is a necessary condition to ensure the connection intensity between the entire node set and the subsampled node set. This assumption allows for sparse networks, which is also assumed by \cite{wang2017likelihood}, \cite{hu2020corrected}, and \cite{li2020network}.

Based on the above assumptions, we now discuss the lower bound of the subsample size. The estimation of the community labels relies on the connection information between the entire node set and the selected nodes. To ensure the statistical accuracy, we consider two constraints on the selected node set. First, the selected node set should contain nodes from all communities in the entire node set. Specifically, denote a set as $\mM_{K}=\{\mS\in [N]: \forall \ k \in [K], \exists \ i \in \mS, z_{i}=k\}$, and then each element in $\mM_{K}$ completely covers $K$ distinct communities. Therefore, we aim to extract a subset of nodes $\mS \in \mM_{K}$. Second, the expected average out-degree based on $A^{s}$ should increase as $N$ grows. Define $d_{N}=E(\sum_iD^{s}_{r,i}/N)$ and then we require $d_{N}=\Omega(\log{N})$. Based on these two constraints, we provide the following proposition.
\begin{proposition}[\textbf{Subsample size}]\label{pro_sampleSize}
Under Assumptions \ref{ass: balance}--\ref{ass: sparsity}, suppose $\mS$ is collected by simple random sampling $n$ nodes from the entire network. If the subsample size $n\ge (\rho_N\alpha_N)^{-1}\log{(KN)},$ then we have $\mS \in \mM_{K}$ and $d_{N}=\Omega(\log{N})$ with probability at least $1-1/N$.
\end{proposition}
\noindent The proof of Proposition \ref{pro_sampleSize} is provided in Appendix B.1 of the Supplementary Material. According to Proposition \ref{pro_sampleSize}, the lower bound of subsample size $n$ depends on three factors: (1) the number of communities $K$; (2) the balance level of community size $\alpha_N$; (3) the network density $\rho_N$. Specifically, if the network contains many communities, subsample size $n$ should be relatively large.
Additionally, the balance level is negatively correlated with subsample size $n$. Under Assumption \ref{ass: balance}, $\alpha_N=\Omega\{(\log{N})^{-1}\}$, the required minimum sample size is $n=\Omega\{(\log{N})^2/\rho_N\}$. In particular, take $\rho_N=(\log{N})^{-1}$, then the sample size is only $n=\Omega\{(\log{N})^3\}$. This implies that, for a moderate sparse network with a limited number of communities and relatively balanced community sizes, the required sample size $n$ can be quite small.

\csubsection{Population Analysis of the SSC Algorithm}

For an observed network, we know that the SSC algorithm is applied to the adjacency matrix. Population analysis implies that we discuss the theoretical properties of SSC based on the SBM probability matrix defined as \eqref{equ_pro}, rather than the adjacency matrix. Following the theoretical analysis techniques of \cite{qin2013regularized} and \cite{lei2015consistency}, we provide the properties of the SSC algorithm based on population analysis.

We introduce more network notations corresponding to the population normalized sub-adjacency matrix as follows. We denote the SBM probability matrix (population adjacency matrix) as
\begin{equation}\label{equ_pro}
\mA=ZBZ^{\top} \in \mR^{N\times N},
\end{equation}
where $Z \in \{0,1\}^{N \times K}$ is a membership matrix, with each row having only one nonzero entry, $Z_{iz_{i}}=1$. Then, the population normalized adjacency matrix is $\mL=\mD^{-1/2}\mA \mD^{-1/2}$, where $\mD$ is the degree matrix of $\mA$. Similarly, under the SSC algorithm, the population sub-adjacency matrix is defined as $\mA^{s} \in \mR^{N\times n}$ and the corresponding population normalized sub-adjacency matrix  as $\mL^{s}=(\mD^{s}_{r})^{-1/2}\mA^{s} (\mD^{s}_c)^{-1/2} \in \mR^{N\times n}$, where $\mD_{r}^{s}=\diag\{(\mD^{s}_{r,i})_{i=1}^{N}\}$ and $\mD^{s}_c=\diag\{(\mD^{s}_{c,j})_{j=1}^{n}\}$, and the entries are $\mD^{s}_{r,i}=\sum_{j=1}^{n} \mA^{s}_{ij}$, $\mD^{s}_{c,j}=\sum_{i=1}^{N} \mA^{s}_{ij}$. Furthermore, the embedding vectors are the largest $K$ left-singular vectors of $\mL^{s}$, denoted by $\wh{\mU}.$

Next, under the SBM, we discuss the block structure of population embedding vectors $\wh{\mU}$ obtained by using the SSC method. The following proposition shows the connection between the membership matrix $Z$ and population embedding vectors $\wh{\mU}$.
\begin{proposition}{\bf (Structure of singular vectors)} \label{pro_community}
Suppose $A$ generated from SBM with $K$ communities, and the subsampled node set $\mS$ is collected by simple random subsampling. Under Assumptions \ref{ass: balance}--\ref{ass: sparsity}, if the sample size $n$ satisfies the condition in Proposition \ref{pro_sampleSize}, there exists a matrix $\mu \in \mR^{K\times K}$ such that $\wh{\mU}=Z\mu$. Furthermore, we have $Z_{i}\mu=Z_{j}\mu  \Longleftrightarrow Z_{i}=Z_{j},$ where $Z\in \mR^{N\times K}$ is a membership matrix and $Z_{i}$ is the $i$th row of $Z$.
\end{proposition}
\noindent The proof of Proposition \ref{pro_community} is provided in Appendix B.2 of the Supplementary Material. According to Proposition \ref{pro_community}, the population embedding vectors have exactly $K$ distinct rows. More importantly, if the $i$th and $j$th rows of $\wh{\mU}$ are equal (i.e., $Z_{i}\mu=Z_{j}\mu$), nodes $i$ and $j$ belong to the same cluster. This is an important conclusion for the SSC algorithm. Recall that, in the SSC algorithm for SBM, k-means clustering is applied to the rows of observed embedding vectors $\wh{{U}}$. Then, under the mild conditions discussed in the next section, one can verify that $\wh{{U}}$ converges to $\wh{\mU}$. Therefore, $\wh{{U}}$ has roughly $K$ distinct rows as well. Applying k-means clustering to $\wh{{U}}$, we can estimate the block membership matrix $Z$. Next, we establish the theoretical properties of SSC empirically.

\csubsection{Consistency of the SSC Algorithm}

Next, we show the convergence of the largest $K$ left-singular vectors of the empirical normalized sub-adjacency matrix. Consequently, the consistency of the SSC algorithm can be established by studying the upper bound of its misclustered rate.

\begin{theorem}[\textbf{Convergence of singular vectors}]\label{the_Uk} Assume $\lambda_{1}\ge \cdots \ge \lambda_{K}$ are the eigenvalues of normalized sub-adjacency matrix  $\mL^{s}$. Let $\delta_{N}=\min_{i}\{\mD^{s}_{r,i}\}$ denote the minimum expected out-degree. If Assumptions \ref{ass: balance}--\ref{ass: sparsity} holds, there exists an orthogonal matrix $\mathscr{O} \in \mR^{K\times K}$ such that
\begin{equation}\label{equ_Uk*}
\|\wh{U}-\wh{\mU}\mathscr{O}\|_{F}  \le \frac{8\sqrt{6}}{\lambda_{K}}\sqrt{\frac{K\log\{4N(N+n)\}}{\delta_{N}}},
\end{equation}
with probability at least $1-1/N$, where $\| \cdot \|_{F}$ denotes the Frobenius norm of a matrix.
\end{theorem}
\noindent
The proof of Theorem \ref{the_Uk} can be found in Appendix B.3 of the Supplementary Material. To illustrate the estimation error bound given in \eqref{equ_Uk*}, we provide the following explanations. First, the error bound is related to $\lambda_{K}.$ According to \cite{rohe2011spectral}, $\lambda_{K}$ is the eigengap of $\mL^{s}$. Moreover, they pointed out that the eigengap cannot be too small. An adequate eigengap ensures that the population embedding vectors can be estimated well. Second, the error bound is lower if minimum degree $\delta_N$ is higher. Considering $\delta_N=\min_{i}\{\mD^{s}_{r,i}\}= O(n\rho_N)=O(\alpha_N^{-1}\log{N}).$ Thus, if $\lambda_{K}$ can be lower bounded by a positive constant, then we have $\|\wh{U}-\wh{\mU}\mathscr{O}\|_{F}=O_P\{(\log{N})^{-1/2}\}.$ Third, the upper bound is also related to the number of clusters, $K$, and the subsample size, $n$. Recall that Proposition \ref{pro_sampleSize} implies that the subsample size $n$ mainly depends on the balance level $\alpha_N$. Therefore, if the number of clusters $K$ is large while the balance level $\alpha_N$ is relatively small, it is difficult to identify the network communities. This conclusion is identical with the one of \cite{lei2015consistency}.

Next, we focus on the clustering error of SSC. First, we give a definition to describe the correctely clustered nodes. Subsequently, according to this definition, we define a misclustered set. Finally, we establish the upper bound of the size of the misclustered nodes in Theorem \ref{the: sbm}.

Let $c_{i} \in \mR^{K}$ denote the observed centroid of the cluster, where the $i$th row of $\wh{U}$ belongs to, for $i=1,2,\cdots, N.$ According to Proposition \ref{pro_community}, $Z_{i}\mu$ is the population centroid corresponding to the $i$th row of $\wh{\mU}$. Hence, if observed centroid $c_i$ is closer to the population centroid $Z_{i}\mu$ than any other population centroid $Z_{j}\mu$ for all $j$ with $j \neq i$, node $i$ is considered correctly clustered. Specifically, we provide the definition of a node to be clustered correctly as follows.
\begin{definition}[\textbf{Clustered correctly}]\label{def_clusteringCorrectly} Node $i$ is clustered correctly, if for any $j \in [N]$, the following inequality holds, $\|c_{i}-Z_{i}\mu\mathscr{O}\| \le \|c_{i}-Z_{j}\mu\mathscr{O}\|$.
\end{definition}
\noindent According to Definition \ref{def_clusteringCorrectly}, we can define the misclustered node set as $\mathscr{R}=\{i: \exists \ j \neq i, \ \|c_{i}-Z_{i}\mu \mathscr{O}\|> \|c_{i}-Z_{j}\mu \mathscr{O}\|\}$. Based on this definition, we then provide the upper bound of the misclustered rate $|\mathscr{R}|/N$ in the next theorem.

\begin{theorem}[\textbf{Consistency results for SBM}]\label{the: sbm} Assume that Assumptions \ref{ass: balance} and \ref{ass: sparsity} are satisfied. Then, there exists a constant $c_{0}$ such that
\begin{equation}\label{equ_misclusteredRate}
\frac{|\mathscr{R}|}{N} \le  \frac{c_{0}K\log\big\{4N(N+n)\big\}}{Nn\rho_N\lambda^{2}_{K}},
\end{equation}
with probability at least $1-1/N$.
\end{theorem}
\noindent The proof is provided in Appendix B.4 of the Supplementary Material. According to Theorem \ref{the: sbm}, we can draw the following conclusions. First, by \eqref{equ_misclusteredRate}, the misclustered rate decreases as the subsample size $n$ increases. Second, for fixed $K$ and $\lambda_{K}$, since $n\rho_N=\Omega(\log{N})$, we have the misclutered rate $|\mathscr{R}|/N=O_P(1/N)$.

Now, we discuss the consistency of the SSC algorithm under the DCSBM. Let $\wh{\mU}^{*} \in \mR^{N\times K}$ denote the row-normalized version of $\wh{\mU}$. Furthermore, let $\wh{\mU}^{*}_{i}$ denote the $i$th row of $\wh{\mU}^{*}$ and $m=\min_{i=1,\cdots, N} \|\wh{\mU}^{*}_{i}\|$ denote the minimum leverage score of $\mL^{s}$ \citep{drineas2012fast}. Then, we have the following theorem for performing SSC algorithm in DCSBMs.
\begin{theorem}[\textbf{Consistency results for DCSBM}]\label{the: dcsbm} Suppose that $A$ is generated from a ${\rm DCSBM}(z,B,\theta)$ with $K$ communities. Under Assumptions \ref{ass: balance} and \ref{ass: sparsity}, if $n$ satisfies the condition in Proposition \ref{pro_sampleSize}, then there exists a constant $c_{1}$ such that
\beqrs
\frac{|\mathscr{R}|}{N} \le \frac{c_{1}K\log\big\{4N(N+n)\big\}}{N n\rho_Nm^2\lambda^{2}_{K}},
\eeqrs
with probability at least $1-1/N$.
\end{theorem}
\noindent The proof of Theorem \ref{the: dcsbm} can be obtained by similar theoretical techniques to prove the results for SBM. Note that $\|\wh{\mU}^{*}\|_{F}^{2}=K$, then the average leverage score $\|\wh{\mU}^{*}_{i}\|$ is $\sqrt{K/N}$. According to Theorem \ref{the: dcsbm}, if $m$ is at the order of $\sqrt{K/N}$, with $\lambda_{K}$ and $K$ fixed, then $|\mathscr{R}|/N$ goes to zero when $n\rho_N$ grows faster than $\log{N}$.

\csection{NUMERICAL STUDIES}

In this section, we present four simulation examples and two real-world datasets to examine the performance of the SSC algorithm.

\csubsection{ Simulation Models and Performance Measurements}

We start with the generation mechanism of the networks. First, for a given $K$, we assume that the community labels are generated by $z_{i} \sim {\rm Multinomial}(\pi)$ independently for all $i=1,\cdots, N$, where $\pi \in (0,1)^{K}$ and $\sum_{k}\pi_{k}=1$.
Second, we define the connectivity matrix as
\beq\label{eq: block}
B=\rho_N\{ (1-\beta) \one_{K}\one_{K}^{\top} + \beta I_{K}\},
\eeq
where $\one_{K} \in \mR^{K}$ is filled with elements 1 and $I_{K} \in \mR^{K \times K}$ is an identity matrix, and the connectivity divergence $\beta \in (0,1)$ measures the community structure strength. For DCSBM, we follow the scenario proposed in \cite{zhao2012consistency}. The parameters $\theta_i$ are independently generated from
\beqrs
P(\theta_i= mx)= P(\theta_i = x)= \frac{1}{2}, \ \text{with} \ x=\frac{2}{m+1},
\eeqrs
which ensures that $E(\theta_i)=1$.

To gauge clustering performance, we consider the misclustered rate, which has been widely used in the investigation of community detection under SBM/DCSBM \citep{gao2018community}. Let $\wh{Z} \in \mR^{N\times K}$ be the estimated membership matrix. Then, the misclustered rate is calculated as
\begin{equation}\label{equ_misclustered}
\mathcal{R}(\wh{Z},Z)=\frac{1}{N} \min_{\mathscr{O} \in \mE_{K}} d(\wh{Z}\mathscr{O}, Z)
\end{equation}
where $\mE_{K}$ is the set of all $K\times K$ permutation matrices and $d(\cdot, \cdot)$ is the distance function of two matrices with the same dimension. It counts the number of different rows between these two matrices, that is, $d(\wh{Z}, Z)=\sum_{i=1}^{N} \bI(\wh{Z}_{i} \neq Z_{i})$. For a reliable evaluation, the random experiments are repeated for $T = 100$ times. All simulations are implemented in Python and run on a Linux server with a 3.60 GHz Intel Core i7-9700K CPU and 16 GB RAM.

\csubsection{Simulation Settings and Results}

We evaluate the performance of the SSC through the following four different examples: the first three simulation settings are designed to investigate the performance of estimating the community labels for the SBM; and the fourth simulation setting is studied for the DCSBM. The details are as follows.

\begin{example}[Consistency of the SSC] \label{exam: consistency}
We let the size of network $N$ grow from 5,000 to 30,000. For each $N$, according to \eqref{eq: block}, we set $\rho_N=0.06$, $\beta=0.5$. We consider the performances evaluated for $K=3, 4, 5,6$, and set $\pi=\one_{K}/K$ for each $K$. Moreover, according to Proposition \ref{pro_sampleSize}, we set $n$ to be $\lceil 15(\log{N})^2\rceil$, where $\lceil x\rceil$ is the smallest integer greater than or equal to a real number $x$. The simulation results are shown in the left panel of Figure \ref{fig: consistency}. As $N$ increases, the misclustered rate of SSC decreases for each number of communities $K$. We observe that the performance of SSC is better for networks with a small number of communities. These results support the theoretical conclusions of Proposition \ref{pro_sampleSize} and Theorem \ref{the: sbm}. We further compare the computational time of the SSC method with that of the SC method under $K=3$ and $N$ varying from 5,000 to 35,000. According to the right penal of Figure \ref{fig: consistency}, as $N$ grows, the computational time of the SC method increases dramatically, while the SSC method is much computationally efficient than the SC approach.

\end{example}

\begin{figure}[]
     \centering
     \begin{subfigure}[b]{0.46\textwidth}
         \centering
         \includegraphics[width=\textwidth]{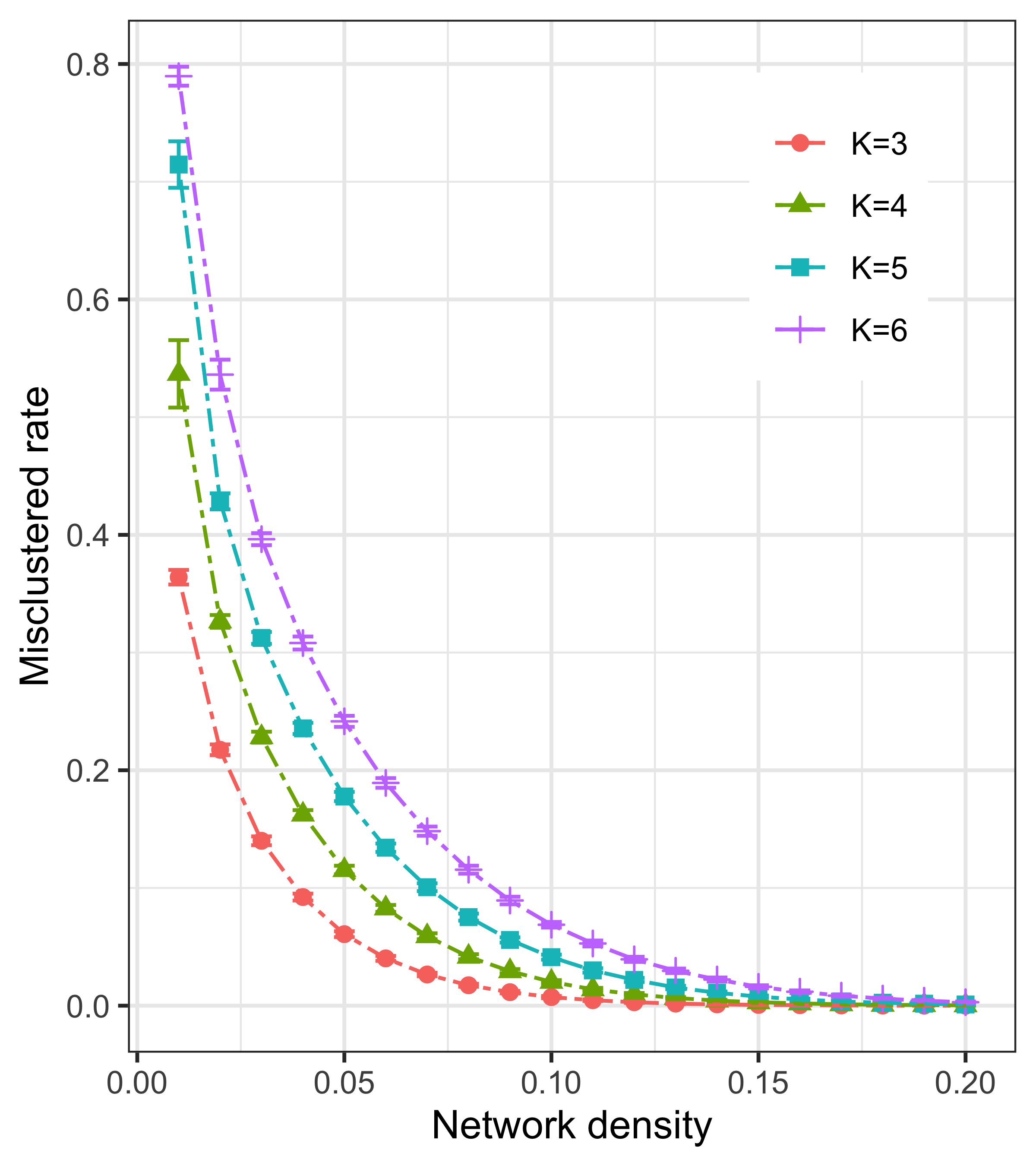}
     \end{subfigure}
     \hfill
     \begin{subfigure}[b]{0.46\textwidth}
         \centering
         \includegraphics[width=\textwidth]{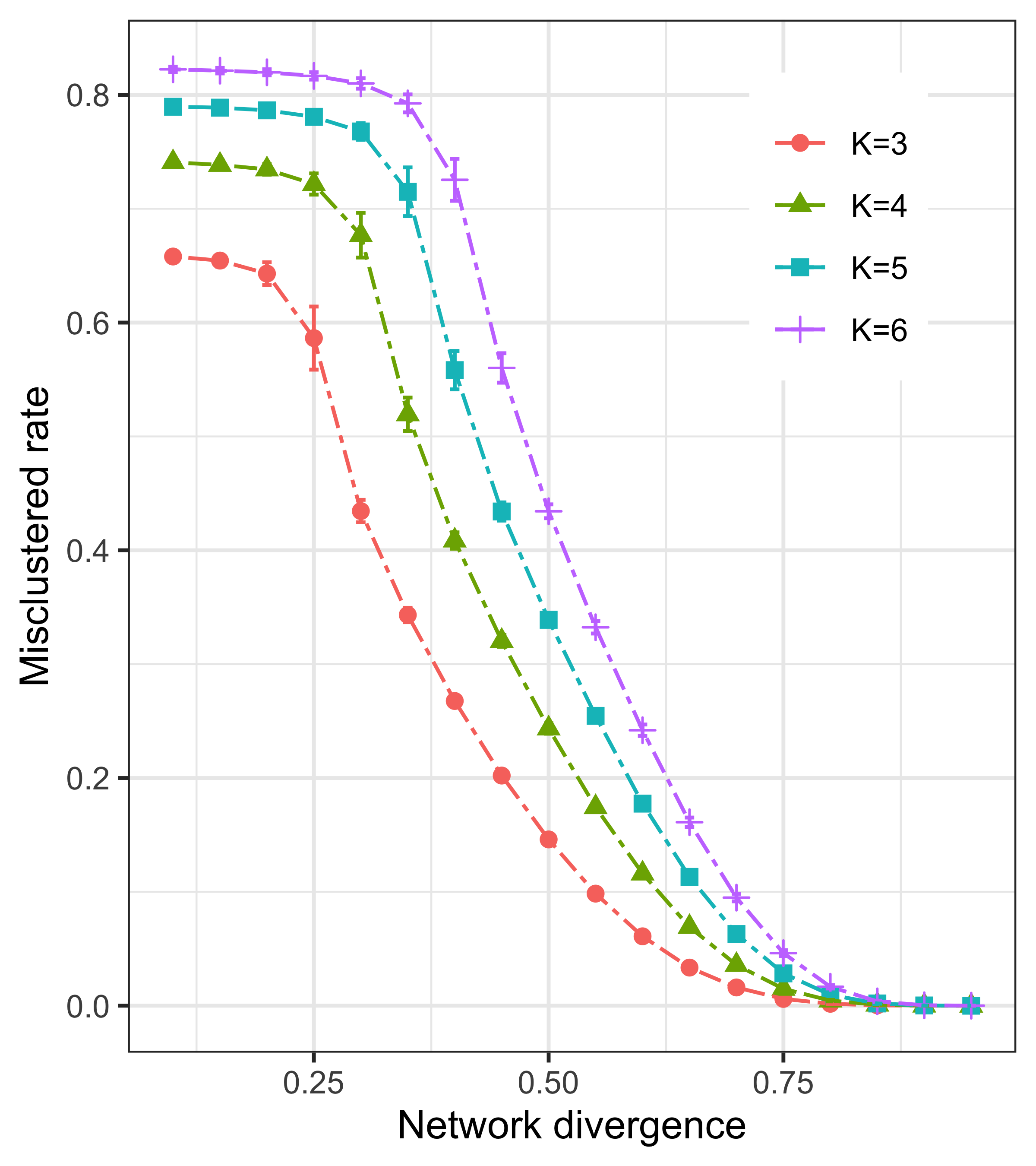}
     \end{subfigure}
      \caption{Simulation results for Example \ref{exam: signal}. The effect of connectivity density $\rho_N$ and connectivity divergence $\beta$ on the misclustered rate of SSC.}
        \label{fig: signal}
\end{figure}

\begin{example}[Effect of signal strength]\label{exam: signal} We fix $N=10,000$ and $n=\lceil 5 (\log{N})^2 \rceil$. Under the SBM, the signal strength of the community structure depends on $\rho_N$ and $\beta$. Here we consider two cases: (1) for $\beta=0.6$, we set $\rho_N$ to increase from 0.01 to 0.20; (2) fixing $\rho_N=0.05$, we vary $\beta$ from 0.05 to 0.95. We also consider the performances evaluated for $K=3, 4, 5, 6$, and set $\pi=\one_{K}/K$ for each $K$. The simulation results are presented in Figure \ref{fig: signal}. According to the left panel of Figure \ref{fig: signal}, as network density $\rho_N$ grows, the misclustered rate rapidly decreases to 0.0 for all settings $K=3, 4, 5, 6$. As shown in the right panel of Figure \ref{fig: signal}, for each setting $K$, when the network divergence $\beta$ increases, the misclustered rate drops to 0.0.
\end{example}

\begin{figure}[]
     \centering
         \includegraphics[width=3.5in]{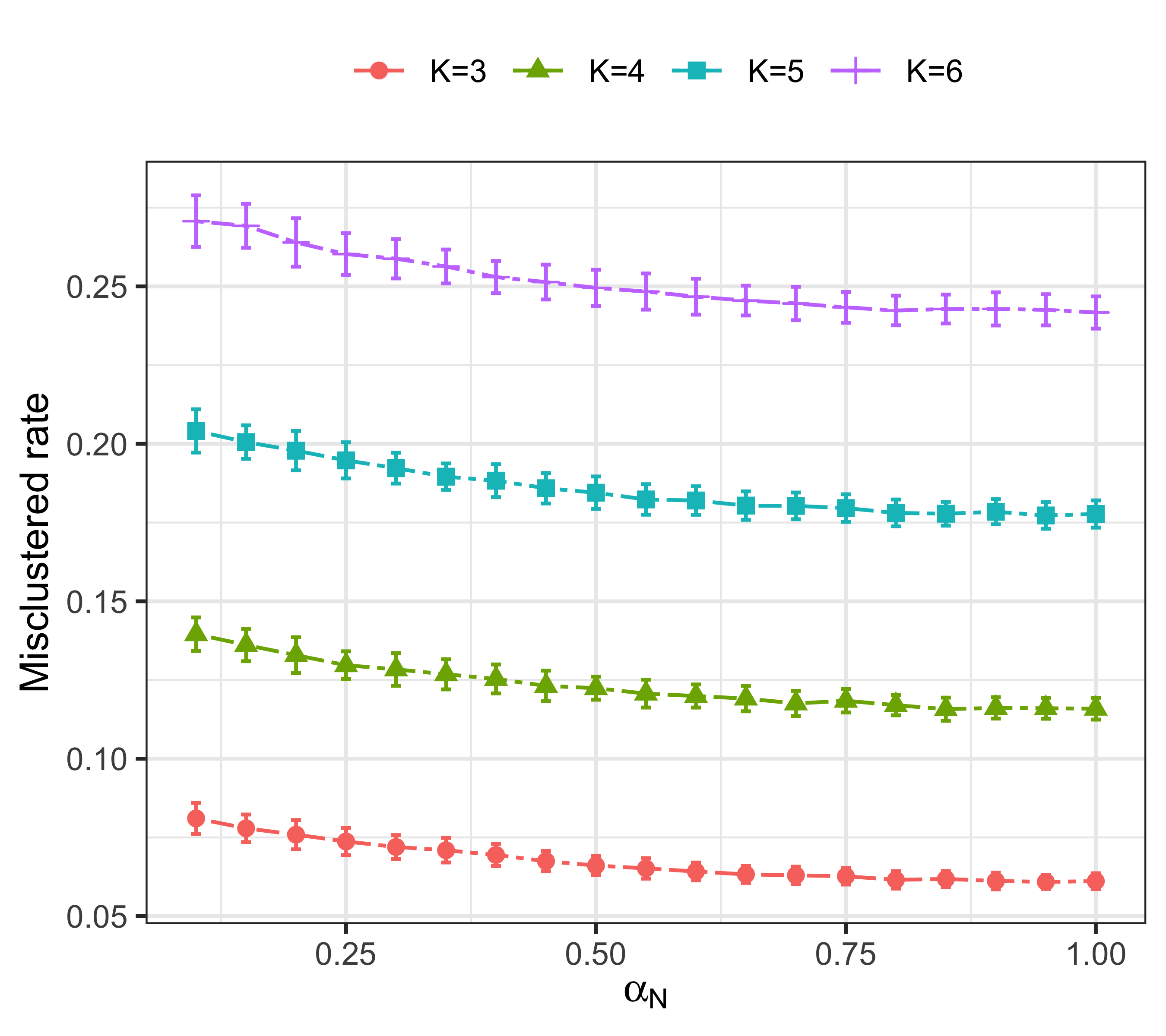}
      \caption{Simulation results for Example \ref{exam: balance}. For each number of communities $K$, the average misclustered rate of SSC decreases as $\alpha_N$ grows.}\label{fig: balance}
\end{figure}

\begin{example}[Effect of balance level]\label{exam: balance} We fix the network size $N=10,000$, $\rho_N=0.05$, $\beta=0.6$, and $n=\lceil 5 (\log{N})^2 \rceil$. To reflect the balance of the network cluster, we set $\pi_{k}$ as
$$\pi_{k}= \frac{1}{K}+ \left( k- \frac{K+1}{2}\right)\frac{(1-\alpha_N)}{K(K-1)}.$$
Moreover, let $\alpha_N$ increase from 0 to 1. Note that a larger $\alpha_N$ implies a higher balance level of the community size. According to Figure \ref{fig: balance}, for each number of communities $K$, we observe that the misclustered rate drops to a lower value as the balance level $\alpha_N$ grows.
\end{example}

\begin{figure}[]
     \centering
     \begin{subfigure}[b]{0.32\textwidth}
         \centering
         \includegraphics[width=\textwidth]{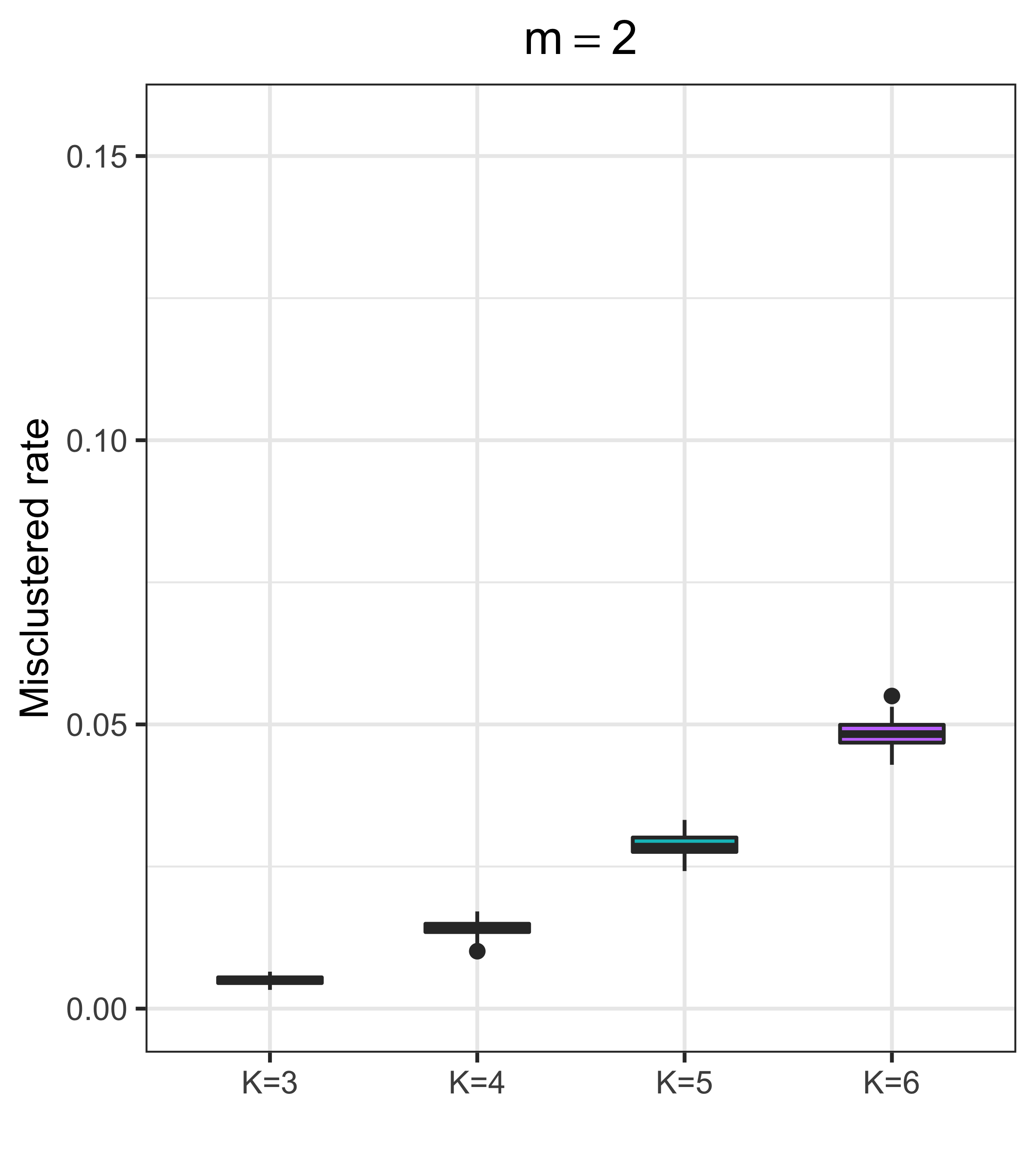}
     \end{subfigure}
     \hfill
     \begin{subfigure}[b]{0.32\textwidth}
         \centering
         \includegraphics[width=\textwidth]{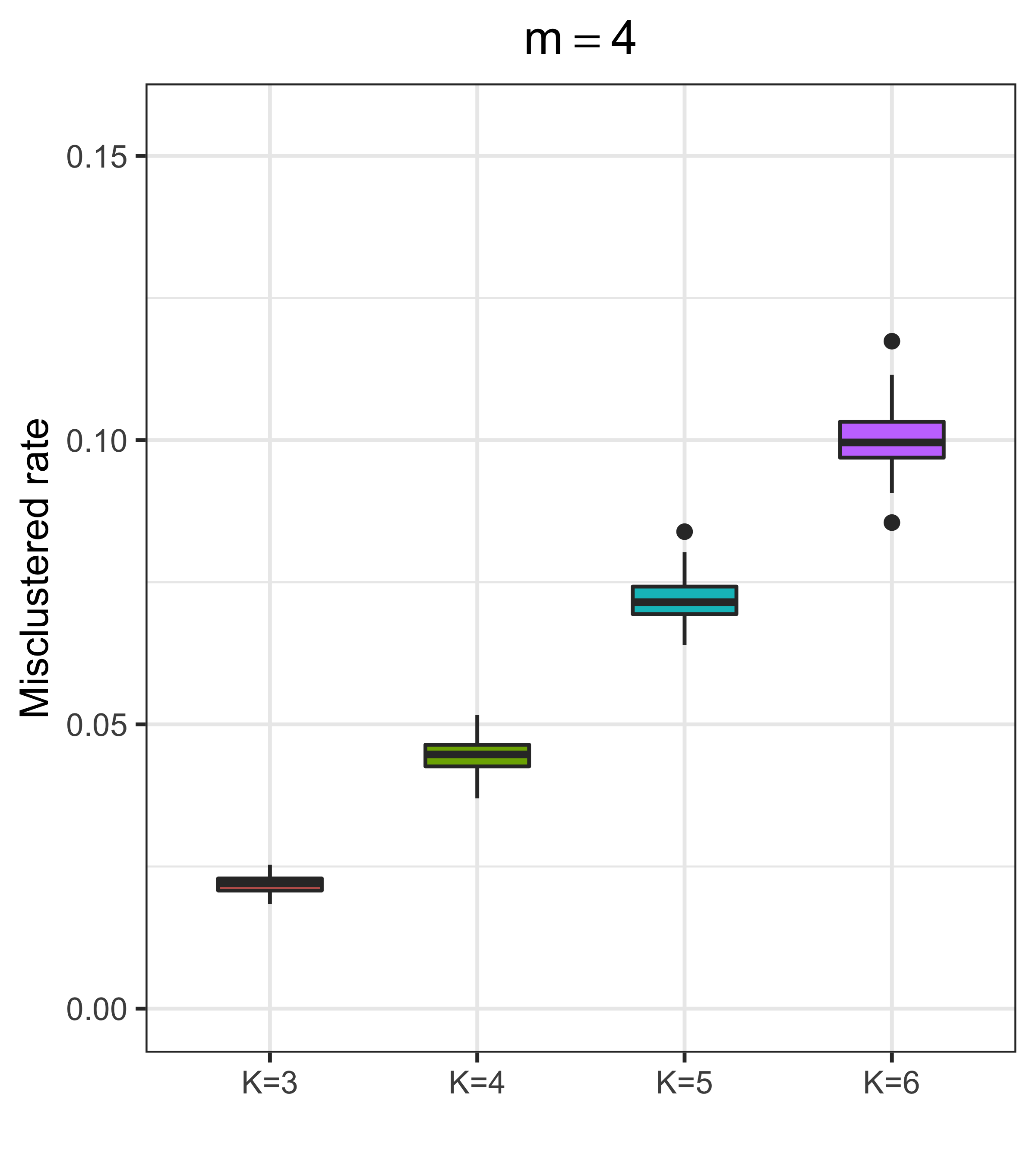}
     \end{subfigure}
        \hfill
     \begin{subfigure}[b]{0.32\textwidth}
         \centering
         \includegraphics[width=\textwidth]{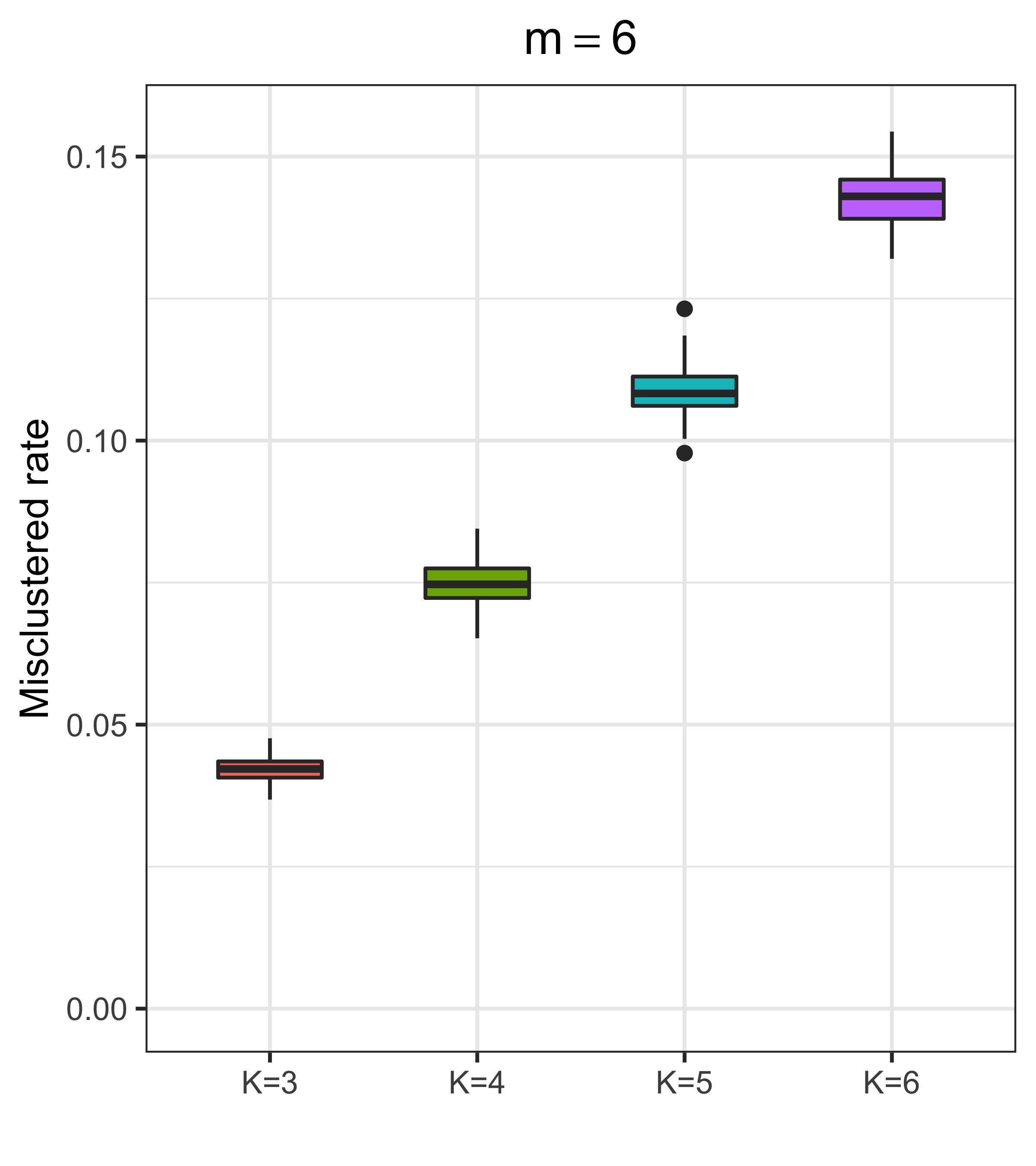}
     \end{subfigure}
      \caption{Simulation results for Example \ref{exam: degree}. For each number of communities, the misclustered rate decreases as the degree heterogeneity $m$ grows from 2 to 6.}\label{fig: degree}
\end{figure}

\begin{example}[Effect of degree heterogeneity]\label{exam: degree} We consider $m$ to be 2, 4, 6, accordingly, where a larger $m$ represents a higher degree heterogeneity effect. Moreover, we set the network size $N=10,000$, $\rho_N=0.05$, $\beta=0.6$, and $n=\lceil 15(\log{N})^2\rceil$. We also consider the performances evaluated for $K=3,4,5,6$, and set $\pi=\one_{K}/K$ for each $K$. The experiment results are shown in Figure \ref{fig: degree}, as $m$ decreases from 6 to 2, the misclustered rate and standard deviation both decrease.
\end{example}

Based on the numerical performance of SSC on these simulation studies, our proposed method is efficient and robust for analyzing large-scale networks.

\csubsection{Real Data Analysis}

\textbf{A Sina Weibo dataset.} We now evaluate the SSC using a dataset collected from Sina Weibo ({\it www.weibo.com}), one of the largest Twitter-type social networks in China. Each node is a user, and an edge exists if there is a follower--followee relationship between two users. An undirected network is constructed based on this dataset. Given any two users, $i$ and $j$, the corresponding element of adjacency matrix $A_{ij}$ is set to 1 if there is at least one edge between two users. This network has $N= 9,980$ nodes and the network density is $1.3\%$.

It is noteworthy that, since the true community labels are unknown in the real data analysis, we take the partition results of spectral clustering as the true labels to evaluate the performance of SSC. First, we use the corrected Bayesian information criterion discussed in Remark \ref{remark: determine} to estimate the number of communities, and we obtain $\wh{K}=3$. Then, under the DCSBM framework, we apply the SSC algorithm with the subsample size $n=\{15(\log{N})^2\}\rceil=1271$ for this network.

\begin{figure}[tbph]
  \centering
  \includegraphics[width=3.5 in]{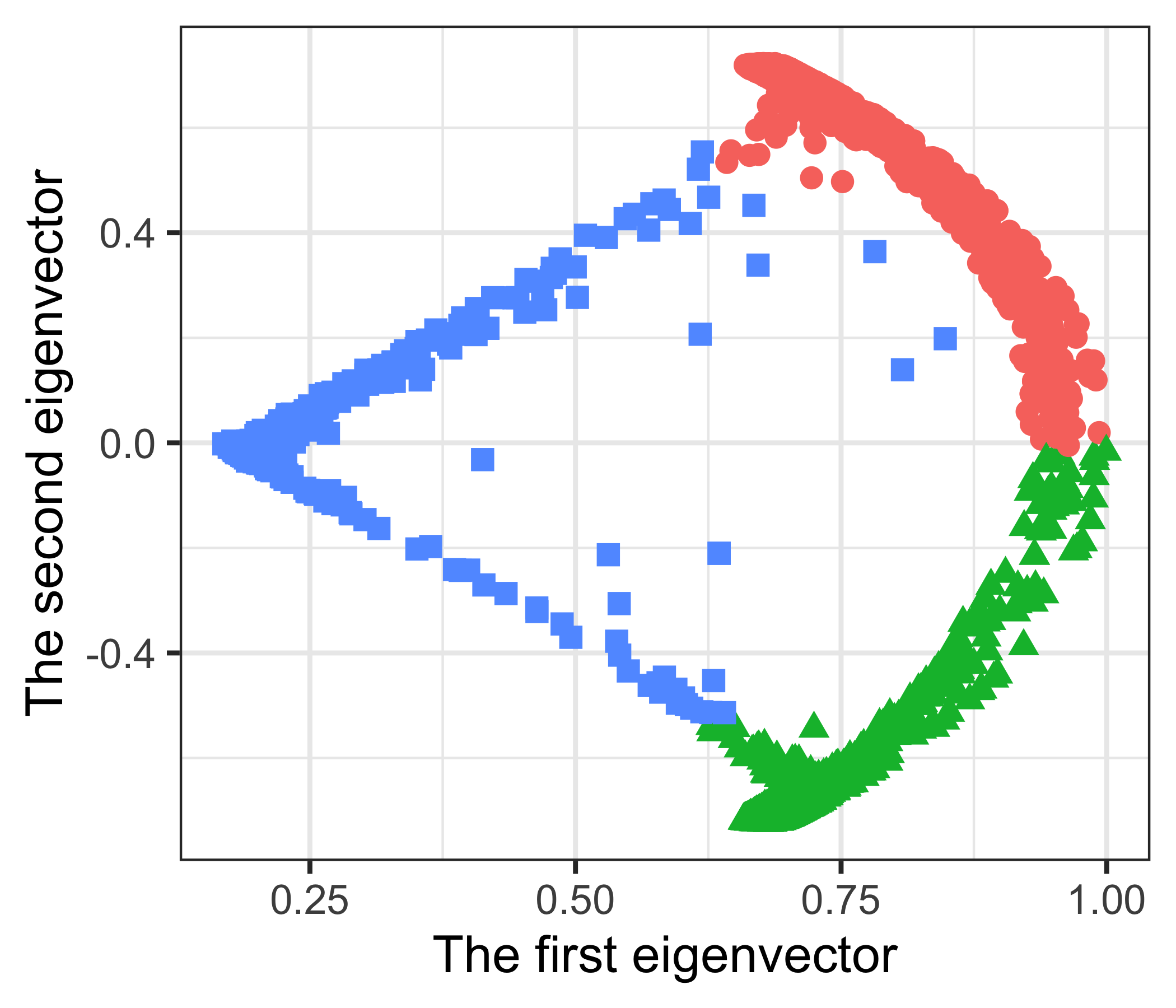}
  \caption{\small Largest two eigenvectors corresponding to the network normalized adjacency matrix $L$. The x-axis provides the first largest eigenvector and the y-axis the second largest eigenvector values. The type of scattered points represents community label. }\label{fig: eigenvectors}
\end{figure}

We demonstrate the clustering results of both the spectral clustering and SSC methods. First, based on the spectral clustering algorithm, Figure \ref{fig: eigenvectors} shows that the embedding vectors of network nodes have a clear community structure. This indicates that spectral clustering is feasible for identifying the community labels of this network. Second, the misclustered rate of the SSC is 0.02 and its computational time is 0.52s, while the computational time of spectral clustering based on the entire network is 2.92s. Namely, SSC has comparable clustering accuracy to SC. For larger networks, the SSC method is feasible even using a personal computer to obtain the cluster labels for the entire network.

\begin{figure}[]
     \centering
     \begin{subfigure}[b]{0.32\textwidth}
         \centering
         \includegraphics[width=\textwidth]{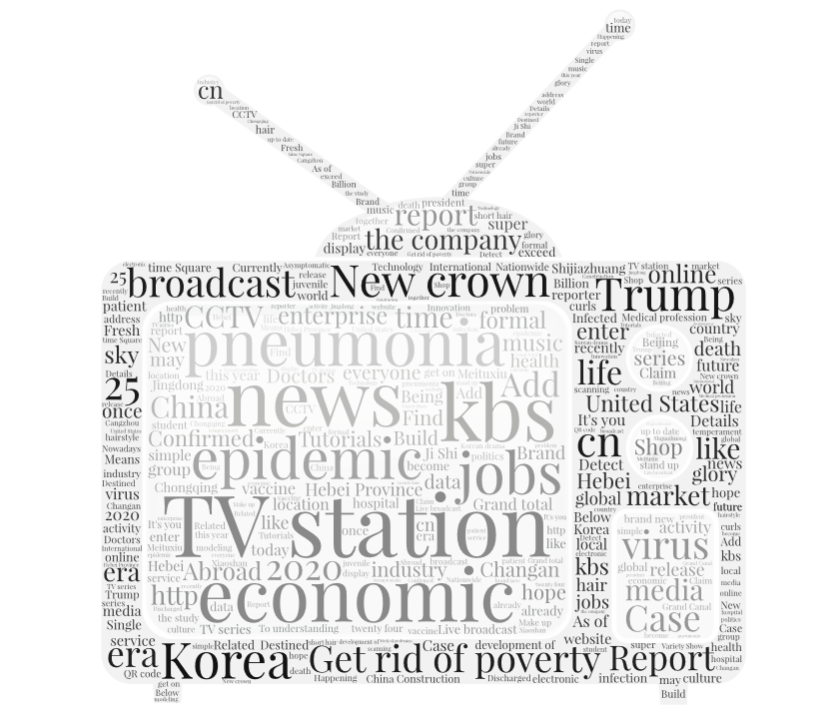}
     \end{subfigure}
     \hfill
     \begin{subfigure}[b]{0.32\textwidth}
         \centering
         \includegraphics[width=\textwidth]{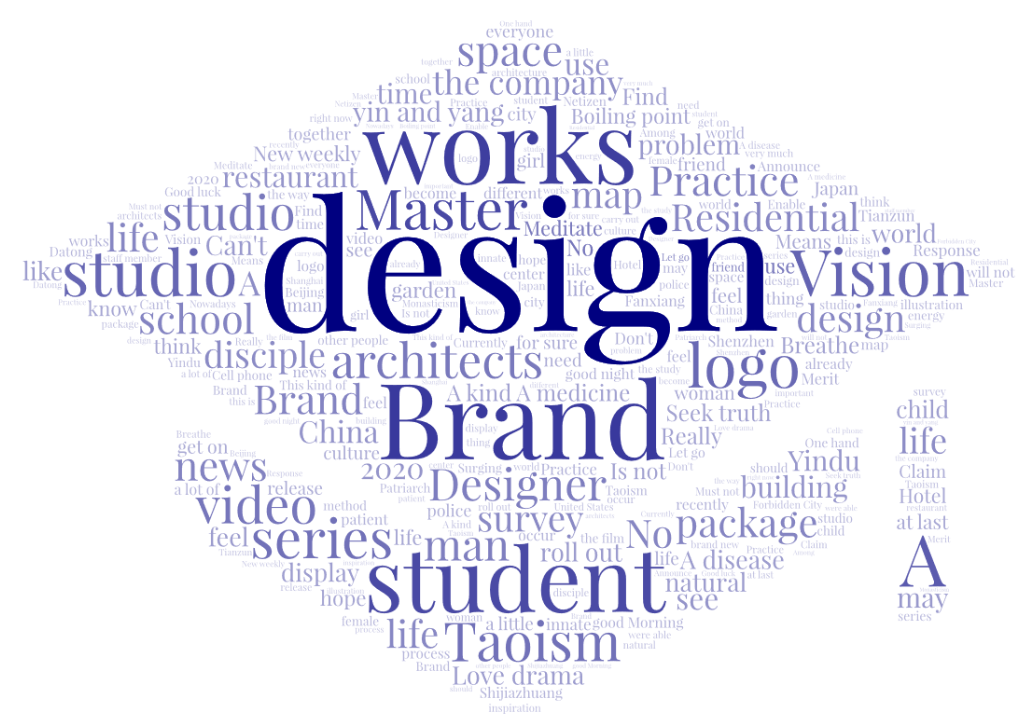}
     \end{subfigure}
        \hfill
     \begin{subfigure}[b]{0.32\textwidth}
         \centering
         \includegraphics[width=\textwidth]{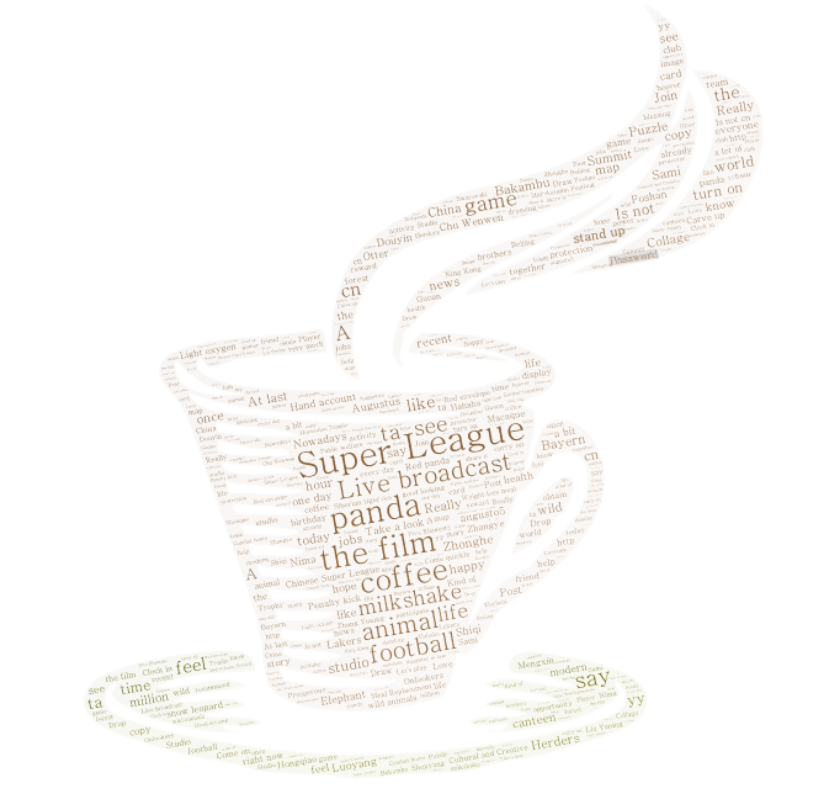}
     \end{subfigure}
  \caption{\small Word clouds of the social network communities discovered by the SSC algorithm under the DCSBM framework. The shapes of word clouds represent the community label and the sizes of the words their frequencies. The left panel includes representative words related to news. The middle panel implies that the users are engaged in work about fashion and art. The right panel shows that the users in this community are used to sharing their leisure habits.}\label{fig: community}
\end{figure}

Furthermore, to explore more interesting information about the community structure obtained by the SSC algorithm, we analyze the text information posted by the users of this social network. As shown in Figure \ref{fig: community}, we plot three {\it word clouds} to depict the representative keywords of each community. The keywords of each community are generated by gathering the Sina Weibo content posted by users. The sizes of the keywords reflect word frequency, and the shapes of word clouds are related to high-frequency words. There are clear differences between the word clouds. Based on these word clouds, the members in different communities have distinct interests and play different roles in the network. The users assigned to the first cluster are highly concerned about the news, the high-frequency keywords of their blogs being ``TV station'' and ``news.'' The second cluster includes users who are interested in fashion and art, their high frequency keywords being ``design,'' ``brand,'' and ``works.'' The third cluster is a group with a wide range of hobbies, and it has high-frequency keywords such as ``league,'' ``film,'' and ``coffee.''

\textbf{ Mnist dataset.} We further evaluate the SSC method on Mnist dataset, which is a large dataset of handwritten digits \citep{lecun1995comparison}. For Mnist dataset, we first select the digit images with labels 0, 1, 3, 5, or 8, and we treat these images as network nodes. Then, the network is generated by connecting each node to its 1000-nearest neighbors, where the distance measure is Euclidean distance. As a result, the network has 30,068 nodes and 5 well-defined communities, and the network density is 4.6\%. Then, under DCSBM framework, we use SC and SSC to identifying the community labels for this network, respectively. For SSC, the subsample size is $n= \lceil \{15(\log{N})^2\}\rceil=1594$. In this network, the misclustered rate of SC and SSC are both 0.537, while the computational time of SC is 94.21 seconds and that of SSC is only 2.75 seconds. In this way, the computational time of the SSC method is only 2.9\% of the spectral clustering based on the whole network.

\csection{CONCLUDING REMARKS}

In this study, we present an SSC algorithm to identify the community structure of large networks. Theoretically, we investigate the subsample size for the SSC method and establish the statistical properties of the clustering results. Specifically, the computational complexity of the SSC algorithm can be reduced to $O(Nn)$, where $n$ is the subsample size, which can be as low as $\Omega\{(\log{N})^{3}\}$. Consequently, the SSC method makes community detection for large-scale networks applicable under limited computational resources. Extensive simulation studies and real data analysis demonstrate the statistical accuracy and computational advantages of the proposed method.

The idea of the paper can be extended to research on other network data with more complex relationships, such as bipartite and multiple networks, which we are currently investigating. Additionally, here we study the subsampling method, which only needs to select the node set once. Another interesting issue in future research is to develop a multi-step subsampling method which can extract richer network structure information.

\newpage


\bibliographystyle{asa}
\bibliography{reference}
\end{CJK}

\end{document}